\documentclass[aps, prb, reprint]{revtex4-1}

\usepackage[pdftex]{graphicx}
\usepackage{hyperref}
\usepackage{amsmath}
\usepackage{epstopdf}
\usepackage{times}

\usepackage{color}

\def\Ef{$E_{\rm F}$}
\def\Kpoint{${\bf K}$}
\def\GammaK{${\bf \Gamma}{\bf K}$}
\def\GammaKM{${\bf \Gamma}{\bf K}{\bf M}$}
\def\GammaM{${\bf \Gamma}{\bf M}$}
\def\hbn{\textit{h}-BN}
\def\mlhbn{ML-\textit{h}-BN}
\def\degc{$^{\circ}{\rm C}$}

\begin{document}

\title{Quasi-freestanding and single-atom thick layer of hexagonal boron nitride as a substrate for graphene synthesis}

\author{D.~Usachov}
\email[]{usachov.d@googlemail.com}
\affiliation{St. Petersburg State University, St. Petersburg, 198504, Russia}

\author{D.~Haberer}
\affiliation{IFW Dresden, P.O. Box 270116, D-01171 Dresden, Germany}

\author{A.~Gr\"uneis}
\affiliation{IFW Dresden, P.O. Box 270116, D-01171 Dresden, Germany}
\affiliation{Faculty of Physics, Vienna University, Strudlhofgasse 4, 1090 Wien, Austria}

\author{H.~Sachdev}
\affiliation{Anorganische Chemie 8.11, Universit\"at des Saarlandes, 66041 Saarbr\"ucken, Germany}

\author{A.~B.~Preobrajenski}
\affiliation{MAX-lab, Lund University, Box 118, 22100 Lund, Sweden}

\author{V.~K.~Adamchuk}
\affiliation{St. Petersburg State University, St. Petersburg, 198504, Russia}

\author{C.~Laubschat}
\affiliation{Institute of Solid State Physics, Dresden University of Technology, D-01062 Dresden, Germany}

\author{D.~V.~Vyalikh}
\affiliation{St. Petersburg State University, St. Petersburg, 198504, Russia}
\affiliation{Institute of Solid State Physics, Dresden University of Technology, D-01062 Dresden, Germany}

\begin{abstract}
We demonstrate that freeing a single-atom thick layer of hexagonal boron nitride (\hbn) from tight chemical bonding to a Ni(111) thin film grown on a W(110) substrate can be achieved by intercalation of Au atoms into the interface. This process has been systematically investigated using angle-resolved photoemission spectroscopy, X-ray photoemission and absorption techniques. It has been demonstrated that the transition of the {\hbn} layer from the ``rigid'' into the ``quasi-freestanding'' state is accompanied by a change of its lattice constant. Using chemical vapor deposition, graphene has been successfully synthesized on the insulating, quasi-freestanding {\hbn} monolayer. We anticipate that the \textit{in situ} synthesized weakly interacting graphene/{\hbn} double layered system could be further developed for technological applications and may provide perspectives for further inquiry into the unusual electronic properties of graphene.
\end{abstract}

\maketitle

\section{Introduction}

The outstanding electronic properties of graphene, a monolayer of carbon atoms tightly packed into a two-dimensional honeycomb lattice, put it at the forefront of current research efforts. The highly unusual nature of charge carriers in graphene which behave as massless Dirac fermions with exceedingly high mobility have raised high expectations regarding its future use in a variety of devices fabricated at the nanometer scale.\cite{Novoselov_Nat_2005, Morozov_PRL_2008} The ability to form structures like nanoribbons \cite{Kosynkin_Nat_2009} and nanomeshes with punched holes \cite{Bai_NNano_2010} also attracts huge attention since the symmetry breaking leads to opening of a band gap which is essential for the operation of transistors and diodes. On the other hand, nanopatterned graphene substrates formed on lattice-mismatched transition metal surfaces \cite{Preobraj_PRB_2008, Usachov_2008_PRB} are very promising as templates for nanoparticles.\cite{Diaye_PRL_2006} Despite a variety of unique properties of graphene, the challenge how to bring graphene-based devices into real life use remains. The development of a reliable technological protocol for effective, large-scale synthesis of high-quality graphene flakes on insulating substrates is one of the main long-term goals in graphene technology.

Among a variety of technologically interesting insulating substrates, hexagonal boron nitride (\hbn) is a promising candidate for synthesis of graphene. Similarly to graphite, {\hbn} is a layered material, composed of weakly interacting monolayers. The graphene and the {\hbn} monolayer (ML) almost perfectly fit to each other. They both have honeycomb crystal structure determined by the $sp^2$ orbital hybridization and exhibit rather similar lattice parameters. The ionicity of the B--N bonds makes {\hbn} a wide bandgap insulator,\cite{Watanabe_NM_2004, Rubio_PRB_1994} while graphene is a zero-gap semiconductor. Recent calculations of the electronic structure of graphene on {\hbn} demonstrated that the graphene/{\hbn} system possesses an energy gap of $\sim$53~meV.\cite{Giovannetti_BN_PRB_2007} However, it is possible to tune the width of the gap systematically from 0 to about 130~meV by applying an external electric field.\cite{Slawinska_PRB_2010} The presence of the gap is essential for achieving a high on/off current ratio of the graphene based field effect transistor.\cite{Xia_NanoLett_2010} Thus, the graphene/{\hbn} combination represents an epitaxial semiconductor/insulator system promising for technological use in electronic devices.

\begin{figure*}[htb]
	\centering
\includegraphics[width=6in]{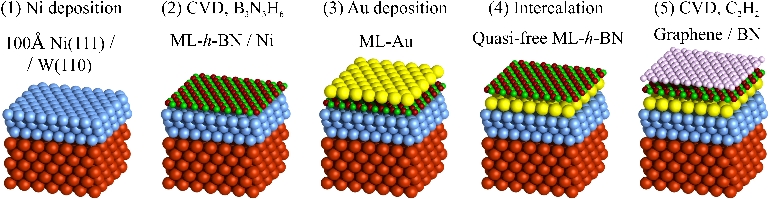}\\
\caption{The procedure for the graphene synthesis on a weakly bonded monolayer of {\hbn}.}
  \label{fig:technology}
\end{figure*}

In the present work we report a route towards large-scale synthesis of graphene on top of the nearly freestanding, single-atom thick, insulating hexagonal boron nitride. A particular issue is the exploitation of structural and electronic properties of two-dimensional crystals for creating multi-layered systems with the desirable characteristics. The electron band structure of the system has been monitored by angle-resolved photoelectron spectroscopy (ARPES) in order to follow its evolution over the different stages of the growth process. Due to the insulating nature of bulk {\hbn}, ARPES cannot be directly applied to graphene/{\hbn} system. However, the feasibility of ARPES experiments on a single-atom thick {\hbn} on Ni(111) has been successfully demonstrated.\cite{Oshima_BN_PRB_1996} We show that the intercalation technique, previously developed for the graphene/Ni interface, is applicable to {\mlhbn} as well and allows to liberate it from the tight chemical interaction with Ni, which is important for the graphene/{\mlhbn} synthesis.

Being a two-dimensional crystal similar to graphene, monolayer {\hbn} is a very attractive object and many research efforts have been undertaken recently to explore its morphology and properties on different metallic substrates.\cite{Oshima_BN_PRL_1995, Preobrajenski_CPL_2007, Preobrajenski_PRB_2008} If combined with strong chemisorption, the monolayer can be strongly corrugated forming a so-called {\hbn} nanomesh, like on Rh(111) \cite{Laskowski_PRB_2010} and Ru(0001).\cite{Goriachko_L_2007, Laskowski_PRB_2010} Among a variety of metal substrates used for synthesis of {\hbn} films, particular attention is paid to the Ni(111) surface, since {\hbn} is almost perfectly lattice matched to it and grows easily.\cite{Oshima_BN_PRB_1995, Preobrajenski_PRB_2004}

However, the graphene synthesis on {\mlhbn}/Ni(111) leads to a misorientation of the flakes, thus preventing a reliable ARPES study of its electronic properties near the Fermi level at the {\Kpoint} point.\cite{Oshima_BN_PRB_1996} In order to obtain uniform in-plane orientation of graphene flakes it is of a great importance to match the lattice constants of graphene and {\mlhbn} as much as possible. The in-plane lattice parameter of bulk {\hbn} ($2.50$~\AA\cite{Paszkowicz_APA_2002}) differs from that of graphite ($2.46$~\AA\cite{Dresselhaus_2002}) by nearly $1.6$\%, but for {\mlhbn} this value may be different, depending on the mismatch and interaction with the substrate. The interaction between {\mlhbn} and Ni was shown to be rather strong.\cite{Preobrajenski_PRB_2004} Using thin Ni films on W(110) as a substrate for {\hbn} growth can induce additional strain into the {\hbn} lattice, because Ni itself is known to grow on W(110) in a stressed fashion.\cite{Sander_1988_PRB} Moreover, this strain will depend on the Ni film thickness. Therefore, releasing the {\mlhbn} from strong interaction with Ni may improve the subsequent growth of graphene by reducing the strain in the {\mlhbn}. To this end we deposited and subsequently intercalated Au atoms into the {\mlhbn}/Ni(111) interface in the same way as it has been done recently for graphene/Ni(111) to bring it into a ``quasi-freestanding'' state.\cite{Shikin_2000_PRB, Varykhalov_2008_PRL} We believe that the well elaborated approach for \textit{in situ} synthesis of weakly interacting graphene on the insulating, single-atom thick layer of quasi-freestanding {\hbn} could open perspectives for further studies of the exotic properties of massless Dirac fermions in graphene.

\begin{figure}
	\centering
\includegraphics[width=2.5in]{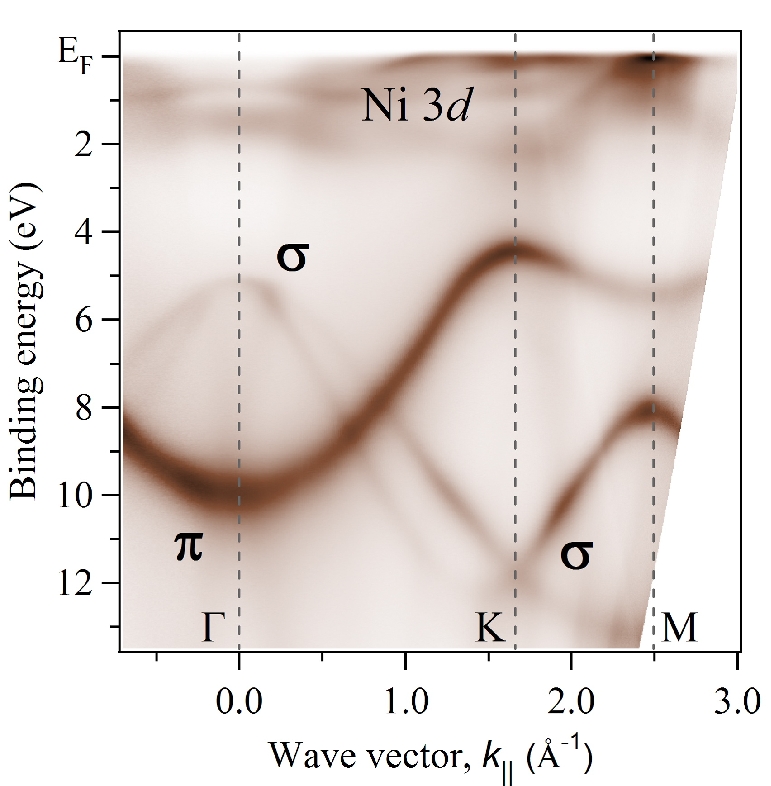}
\caption{The electron energy band structure of the {\mlhbn}/Ni(111)/W(110) system measured by ARPES.}
  \label{fig:BN_GK}
\end{figure}

\begin{figure}
	\centering
(a)\includegraphics[width=1.5in]{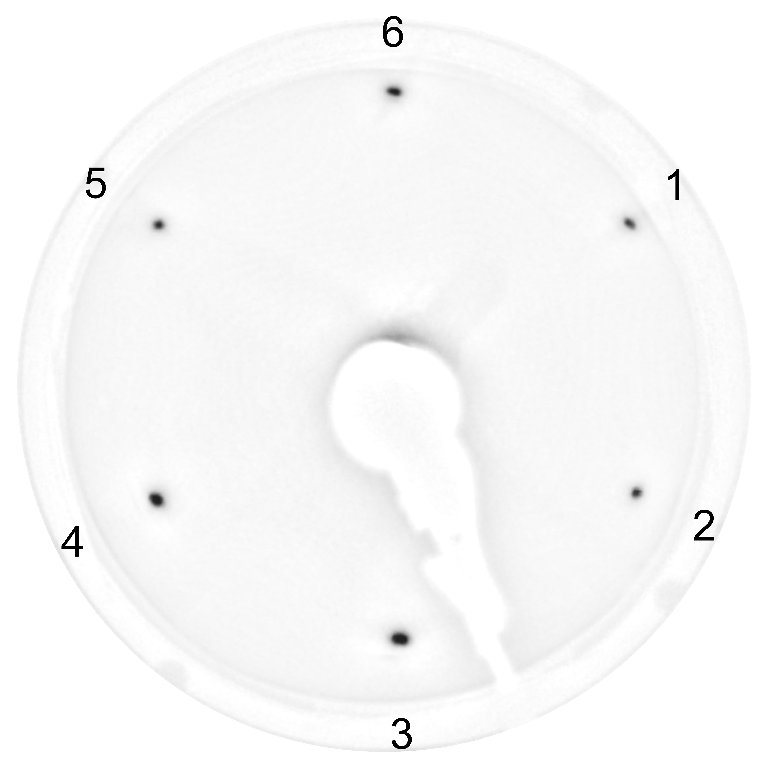} (b)\includegraphics[width=1.5in]{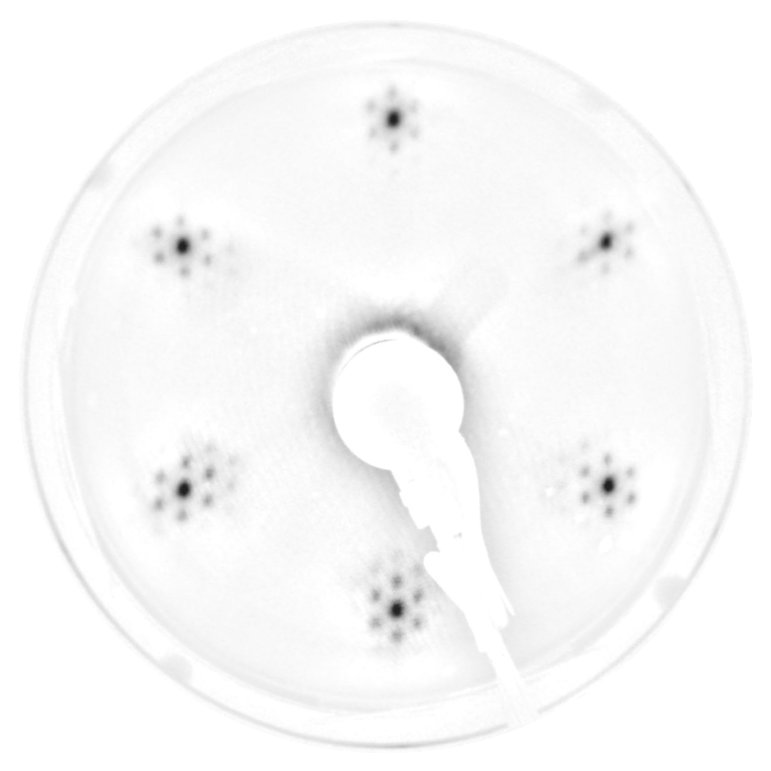}
\caption{The LEED patterns of (a) {\mlhbn}/Ni(111)/W(110) and (b) {\mlhbn}/ML Au/Ni(111)/W(110).}
  \label{fig:LEED}
\end{figure}

\section{Experiment}

The procedure for the \textit{in situ} synthesis of graphene on {\mlhbn} is shown schematically in Fig. \ref{fig:technology}. In the beginning, a clean surface of Ni(111) was prepared (step 1). Instead of using bulk single crystal as a substrate, thin ($\sim 100$~\AA) Ni films have been epitaxially grown on W(110) resulting in Ni(111) layers. Using W(110) enables removing Ni from the W(110) by short ``flashes'' and repeating the preparation sequence for verification purposes. Next, a high-quality {\hbn} layer was produced on top of Ni(111) by cracking the borazine vapor (B$_3$N$_3$H$_6$) at a pressure of $10^{-7}$~mbar and a temperature of $750${\degc} for $10$~min (step 2). Borazine was made according to a modified route described in ref. \onlinecite{Sachdev_1995}. The reaction is known to be self-limited to one monolayer. At the next stage the {\mlhbn} film was liberated from a tight chemical interaction with Ni atoms by depositing Au on top (step 3) and its subsequent intercalation (step 4) by annealing the system at $\sim 500${\degc} for 5~min. At the final stage (step 5), a graphene layer was formed by chemical vapor deposition (CVD) of acetylene (C$_2$H$_2$) on the quasi-freestanding {\mlhbn} at a temperature of $\sim 750${\degc} and an acetylene gas pressure of $\sim 3\times 10^{-4}$~mbar for $\sim 90$~min. The exposure has been chosen to obtain graphene coverage of nearly 0.5~ML, which ensures absence of graphene multilayers. Therefore, the measured electronic structure corresponds to a single-atom thick layer of graphene.

The ARPES experiments were performed using a photoelectron spectrometer equipped with a Scienta SES-200 hemispherical electron-energy analyzer and a high-flux He resonance lamp (Gammadata VUV-5010) in combination with a grating monochromator. All ARPES spectra were acquired at room temperature and a photon energy of $40.8$~eV (He~II$\alpha$), with an angular resolution of $0.2^{\circ}$ and a total energy resolution of $\sim 50$~meV. Electron band dispersions were measured along the {\GammaKM} direction of the Brillouin zone (BZ) by varying the polar-emission angle. The X-ray photoelectron spectroscopy (XPS) and near-edge X-ray absorption spectroscopy (NEXAFS) experiments on {\mlhbn} were carried out using linearly polarized radiation at the Russian-German beam line at BESSY~II synchrotron radiation facility.

\section{Results and discussion}

An overview band structure map taken from the {\mlhbn} on Ni(111) is displayed in Fig. \ref{fig:BN_GK}. It shows photoelectron intensity as a function of energy and momentum along the {\GammaKM} direction of the Brillouin zone. Evidently, a sharp and intense structure of both $\pi$ and $\sigma$ bands indicates high structural quality of {\hbn}. The weakly dispersing bands near the Fermi level belong to the Ni~3\textit{d} states slightly modified by the {\hbn} $\pi$ -- Ni~3\textit{d} orbital mixing,\cite{Preobrajenski_PRB_2008} similarly to the case of graphene/Ni(111).\cite{Oshima_1994_PRB, Vyalikh_dynamics_2009, Vyalikh_2008_PRB} Because of the surface sensitivity of ARPES, the appearance of nickel bands indicates a small thickness of the synthesized {\hbn} and the presence of a single $\pi$ band unambiguously confirms its monolayer nature. The corresponding low-energy electron diffraction (LEED) pattern (Fig. \ref{fig:LEED}a) exhibits a sharp $(1\times 1)$ hexagonal structure indicating high crystalline quality and uniform in-plane orientation of {\hbn} with respect to the substrate.

\begin{figure}
	\centering
\includegraphics[width=2.5in]{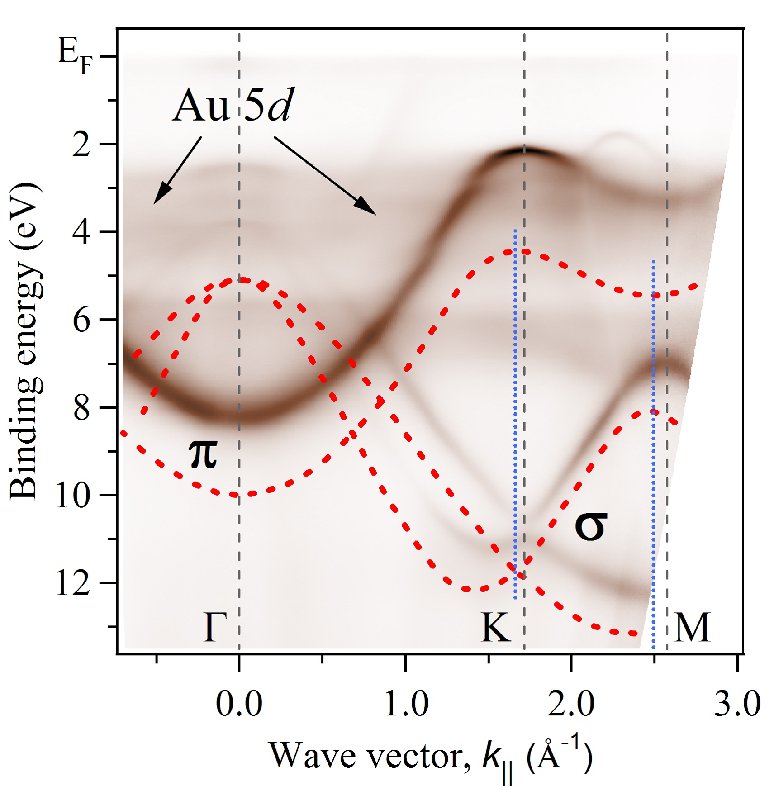}
\caption{The electron energy band structure of the {\mlhbn}/ML-Au/Ni(111)/W(110) system measured by ARPES. The vertical dashed and dotted lines indicate high symmetry points in the BZ of {\mlhbn}/Au and {\mlhbn}/Ni respectively.}
  \label{fig:BN_Au_GK}
\end{figure}

\begin{figure*}
	\centering
\includegraphics[width=7in]{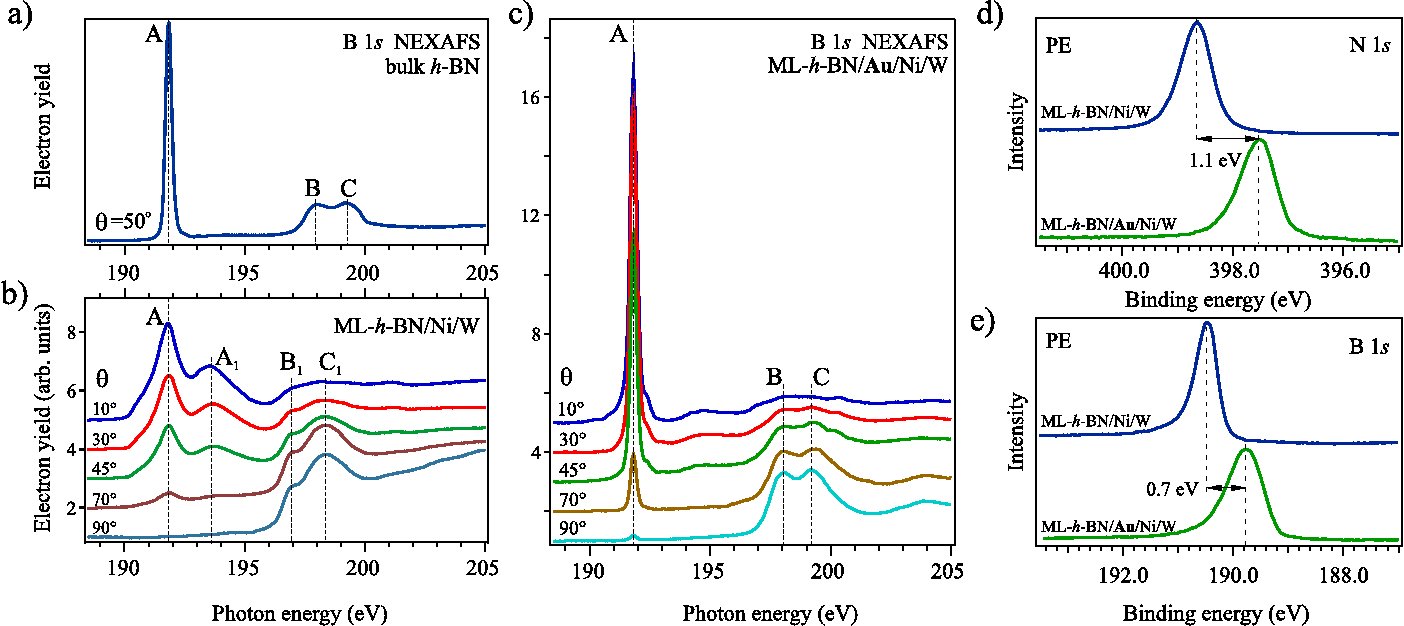}
\caption{NEXAFS spectra of (a) bulk {\hbn} (partial electron yield with retarding voltage 100~V), (b) \mlhbn/Ni(111)/W(110) (total electron yield) (c) \mlhbn/ML-Au/Ni(111)/W(110). Figures (b) and (c) have the same scaling. (d) and (e) shows XPS spectra of the system before and after Au intercalation in the region of (d) N~1\textit{s} ($h\nu=440$~eV) and (e) B~1\textit{s} ($h\nu=230$~eV).}
  \label{fig:spectra}
\end{figure*}

Figure \ref{fig:BN_Au_GK} shows an overview band structure map taken after the Au ML intercalation. For comparison, the $\pi$ and $\sigma$ bands for pristine {\hbn}/Ni(111) are shown by dotted lines. Clearly, the intensity of the Ni~3\textit{d} bands becomes hardly visible, pointing at gold mediated attenuation of the electron emission from the substrate. Neither $\pi$ nor $\sigma$ bands show significant intensity reduction pointing out that the {\mlhbn} remains on top. Both $\pi$ and $\sigma$ bands are shifted towards {\Ef} as it was previously observed for graphene in the quasi-freestanding state.\cite{Shikin_2000_PRB, Varykhalov_2008_PRL} Moreover, the absence of original bands unambiguously demonstrates that the intercalation process is complete and {\mlhbn} has been lifted off the Ni surface. Notably, for the Au intercalated {\mlhbn} the LEED pattern gains an additional set of reflexes (Fig. \ref{fig:LEED}b), indicating the emergence of an approximately $(9\times 9)$ structure due to the mismatch of lattice parameters for {\hbn} and the intercalated Au layer.

The NEXAFS spectra provide convincing evidence for a weakening of the chemical interaction between {\hbn} and Ni atoms upon Au intercalation. Figure \ref{fig:spectra}a shows B K-edge NEXAFS spectrum taken from bulk {\hbn} at the geometry when the angle $\theta$ between surface normal and polarization direction of incident light was set to $50^\circ$. It exhibits the prominent $\pi^{\ast}$-resonance (peak A) and additional $\sigma^{\ast}$-resonance composed mainly of the two features B and C. In contrast to that, the spectral structure of the B K-edge NEXAFS spectra taken from single layer of {\hbn} on Ni(111) looks rather differently in accordance with Ref. \onlinecite{Preobrajenski_PRB_2004}. The corresponding set of spectra recorded at different $\theta$ is displayed in Figure \ref{fig:spectra}b. The $\pi^{\ast}$-resonance is broader now and split mainly into two peaks A and A$_1$ while the $\sigma^{\ast}$-resonance is shifted towards lower energies and noticeably modified in shape. This is known to be a result of orbital mixing with the Ni~3\textit{d} states.\cite{Preobrajenski_PRB_2004} Taking a similar set of spectra after Au intercalation we found that energies for both $\pi^{\ast}$ and $\sigma^{\ast}$ resonances are restored back to the values of bulk {\hbn} (Fig. \ref{fig:spectra}c). Moreover, XPS data taken from the {\hbn}/Ni(111) and the {\hbn}/Au/Ni(111) systems reveal substantial chemical shifts towards {\Ef} for both the N 1\textit{s} (Fig. \ref{fig:spectra}d) and the B 1\textit{s} (Fig. \ref{fig:spectra}e) core levels after Au intercalation.

\begin{figure}
	\centering
\includegraphics[width=3.0in]{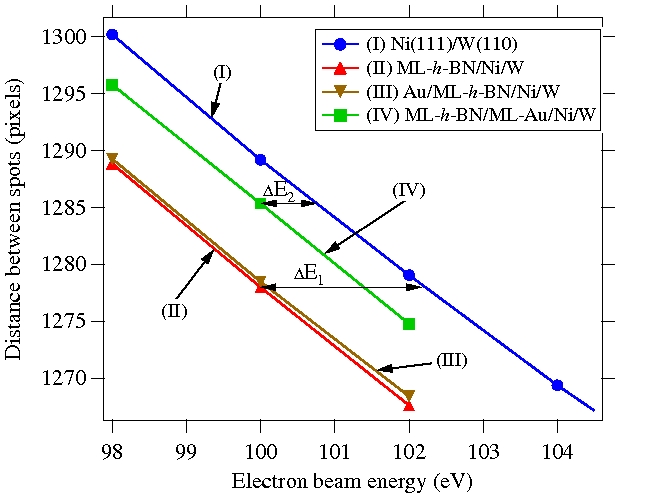}
\caption{The dependence of the distance between the two selected opposite spots on the electron beam energy in the LEED patterns of different systems.}
  \label{fig:LEED_plot}
\end{figure}

All these observations indicate that Au intercalation considerably reduces chemical interaction between Ni and {\mlhbn} leading to a transition of {\hbn} from the chemisorbed to the quasi-free state. We suppose that this makes the {\mlhbn} more flexible for adjusting to graphene. An inspection of the electron band dispersion along {\GammaK} and {\GammaM} (not shown) before and after Au intercalation reveals a slight change of the BZ size. The BZ borders for pristine {\mlhbn}/Ni(111) are shown in Fig. \ref{fig:BN_Au_GK} by short vertically aligned dotted lines. This observation indicates that the liberation of the {\mlhbn} by means of Au intercalation is accompanied by a decrease of its lattice parameter, and supports the idea that the {\mlhbn}/Au/Ni(111) system is more favourable for graphene synthesis than {\mlhbn}/Ni(111). To get quantitative insight into the change of the {\mlhbn} lattice constant upon Au intercalation we recorded LEED patterns keeping the geometry of the LEED setup and parameters of the analysis unchanged. A typical set of LEED images with the assignment for the structural spots is depicted in Figure \ref{fig:LEED}. Fitting the LEED intensity distribution with a two-dimensional Gaussian function, we established coordinates for each spot and analyzed their changes upon Au intercalation. To this end, we determined the distance between opposite spots in the hexagonal LEED patterns for all stages of sample preparation. This is shown in Figure \ref{fig:LEED_plot} by the example of the distance between reflexes 1 and 4 (Fig. \ref{fig:LEED}a) taken for a set of electron beam energies. Apparently, the shift between the different lines corresponds to a change of the lattice parameter. The relative differences between the lattice parameters of {\mlhbn} and the Ni(111) film are determined by the equation:
\begin{equation}
	\delta = \frac{a_{BN} - a_{Ni}}{a_{Ni}} = \sqrt{\frac{E_{Ni}}{E_{BN}}} - 1 \approx \frac{1}{2}\frac{\Delta E}{E_{BN}},
\end{equation}
where $a_{BN}$ -- the {\mlhbn} lattice parameter, $a_{Ni}$ -- the Ni(111) lattice parameter, $E_{Ni}, E_{BN}$ -- the two energies, corresponding to the same distance between spots, and $\Delta E = E_{Ni} - E_{BN}$ is shown in Fig. \ref{fig:LEED_plot}. The results obtained for different spots pairs are summarized in Table \ref{tab:tab1}. The lattice parameter of {\mlhbn}/Ni(111)/W appears to be greater than the surface lattice constant of the Ni(111) film. Note that the values of the lattice constant change obtained from pairs 1--4 and 3--6 are identical. However, they differ from those derived from the pair 2--5, which unambiguously indicates the influence of the W(110) substrate, because the pair 2--5 is aligned along the $[1\overline{1}0]$ direction of W, along which the stress in the Ni film is probably strongest.\cite{Sander_1988_PRB}

\begin{table}[htbp]
	\centering
	\begin{tabular}{|c|c|c|c|}
	\hline
	System	& $\delta\cdot 100\%$	& $\delta\cdot 100\%$	& $\delta\cdot 100\%$	\\
					& (1--4)				& (2--5)				& (3--6)				\\
  \hline
  {\hbn}/Ni/W	& 1.1	& 0.85	& 1.1	\\
  {\hbn}/Au/Ni/W	& 0.35	& 0.25	& 0.35	\\
  \hline
	\end{tabular}
	\caption{The relative differences between the lattice parameters of Ni(111) and {\mlhbn} in different systems.}
	\label{tab:tab1}
\end{table}

The Au deposition does not significantly affect the LEED pattern, which denotes that Au atoms tend to create disordered clusters on the surface. After Au intercalation the lattice parameter of BN shrinks. It should be noted that the LEED evaluated $0.7$\% decrease of the {\mlhbn} lattice parameter upon intercalation is smaller than that estimated from the ARPES data ($\sim 3$\%). This could be explained by a different amount of stress in the Ni films prepared for LEED and ARPES experiments.

\begin{figure}[t]
	\centering
\includegraphics[width=2.5in]{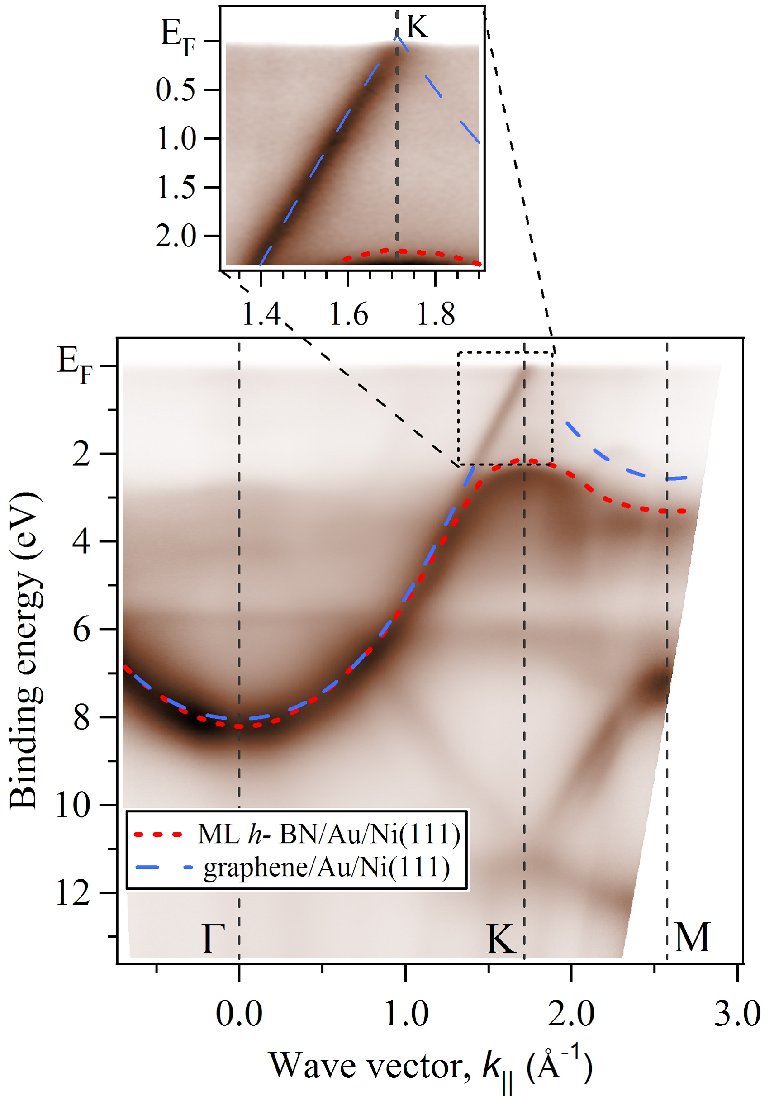}
\caption{The electron energy band structure of the graphene/{\mlhbn}/ML Au/Ni(111)/W(110) system measured by ARPES. The red and blue lines denote the PE maxima of {\mlhbn}/Au and graphene/Au, respectively.}
  \label{fig:Gr_BN_GK}
\end{figure}

Figure \ref{fig:Gr_BN_GK} shows an overview band structure map taken after the CVD of acetylene on the quasi-freestanding {\mlhbn} (step 5). Obviously, a new electron band is detected, which is approaching {\Ef} at the {\Kpoint} point of the BZ. It evidently exhibits linear dispersion at the {\Kpoint} point that allows us to attribute it to the $\pi$ band of graphene. The interesting region at the {\Kpoint} point, framed by the dotted rectangle in Fig. \ref{fig:Gr_BN_GK}, is depicted as an inset. The sharpness of this band and its linear behavior confirm our assumption that the quasi-freestanding {\mlhbn} has a positive effect on the orientation of the graphene layer. The second branch of the Dirac cone is hardly visible due to the geometry of the ARPES experiment. However, comparison of the energy position of the $\pi$ band of graphene/{\hbn}/Au/Ni(111) with that of graphene/Au/Ni(111) suggests that the apex of the Dirac cone is located close to the Fermi level.\cite{Varykhalov_2008_PRL} The existence of the theoretically predicted energy gap \cite{Giovannetti_BN_PRB_2007} is hardly detectable. It could possibly be disclosed by transport measurements, but a thicker {\hbn} film is needed in order to achieve the necessary electrical insulation.

In summary, a single-atom thick layer of hexagonal boron nitride has been grown on a Ni(111) film deposited onto a W(110) substrate. We have shown that the {\hbn} monolayer can be moved to a quasi-freestanding state by means of Au intercalation to the {\mlhbn}/Ni interface. This process is accompanied by the change of the {\mlhbn} electronic structure and the lattice constant as well. Evidently, this structure exhibits enough thermal stability to enable growth of graphene on top of the {\mlhbn}. The structural order of the graphene layer synthesized on the quasi-freestanding {\mlhbn} allows exploring its electronic structure and, in particular, the behavior of the $\pi$ band, which obviously exhibits linear dispersion near the Fermi level. We anticipate that the presented approach will be a further step towards a large-scale synthesis of graphene on insulating substrates for electronic applications.

\begin{acknowledgments}
This work was supported by the DFG Grant No. VY 64/1-1 and RFBR Grants No. 08-03-00410 and No. 10-08-00580. D.~Usachov thanks the DAAD for support within the program ``partnerships with universities in Central and Eastern Europe''. A.~Gr\"uneis acknowledges the DFG project GR 3708/1-1, an APART fellowship from the Austrian Academy of Sciences and a Marie Curie reintegration grant (ECO-GRAPHENE). We acknowledge Roland H\"ubel for technical assistance.
\end{acknowledgments}

\end{document}